\documentclass[useAMS,usenatbib]{almamemo}
\usepackage{amssymb}
\usepackage{graphicx}
\usepackage{mathpazo}
\usepackage{xspace}
\usepackage{booktabs}
\usepackage{prettyref}
\usepackage[pdftex]{hyperref}
\hypersetup{
    bookmarks=true,         
    unicode=false,          
    pdftoolbar=true,        
    pdfmenubar=true,        
    pdffitwindow=false,     
    pdfstartview={FitH},    
    pdfnewwindow=true,      
    colorlinks=true,       
    linkcolor=red,          
    citecolor=blue,        
    filecolor=magenta,      
    urlcolor=magenta           
}

\newrefformat{fig}{\hyperref[{#1}]{Fig.\,\ref*{#1}}}
\newrefformat{sec}{\hyperref[{#1}]{\S \,\ref*{#1}}}
\newrefformat{apdx}{\hyperref[{#1}]{Appendix \,\ref*{#1}}}
\newrefformat{tab}{\hyperref[{#1}]{Tab.\,\ref*{#1}}}
\newrefformat{eqn}{\hyperref[{#1}]{Eq.\,\ref*{#1}}}
\newcommand{\atm}{{\sc atm}\xspace}
\newcommand{\aatm}{{\sc aatm}\xspace}
\newcommand{\htwoo}{H$_2$O\xspace}
\newcommand{\otwo}{O$_2$\xspace}

\newcommand{\param}{$\Delta S_\nu/\Delta c$\xspace}
\newcommand{\tg}{$T_\rmn{g}$\xspace}
\newcommand{\tlr}{$\Gamma_\rmn{T}$\xspace}
\newcommand{\taucso}{$\tau_{225}$\xspace}
\volume{590}
\title[Atmospheric dispersion and phase calibration for ALMA]{ALMA
  Memo 590\\Atmospheric dispersion and the implications for phase calibration}
\author[E.\ I.\ Curtis et al.]{Emily
  I.\ Curtis$^{1,2}$, Bojan Nikolic$^{1,2}$, John S.\ Richer$^{1,2}$ and Juan R.\ Pardo$^{3}$\vspace*{1pt}\\
$^1$Astrophysics Group, Cavendish Laboratory, J.\ J.\ Thomson Avenue, Cambridge, CB3 0HE\\
$^{2}$Kavli Institute for Cosmology, c/o Institute of Astronomy, University of Cambridge, Madingley Road, Cambridge, CB3 0HA\\
$^{3}$Centro de Astrobiolog\'ia, CSIC/INTA, Ctra de Ajalvir, km 4,
28850, Torrej\'on de Ardoz, Madrid, Spain\\
\textup{E-mail:
\href{mailto:e.curtis@mrao.cam.ac.uk}{\nolinkurl{e.curtis@mrao.cam.ac.uk}}
(EIC)}\\
\textup{Software and scripts are available from \url{http://www.mrao.cam.ac.uk/~eic22/ALMAmemos/memos}}}
\begin{document}

\date{10 December 2009}

\pagerange{\pageref{firstpage}--\pageref{lastpage}} \pubyear{2009}

\maketitle

\label{firstpage}

\begin{abstract}

The success of any ALMA phase-calibration strategy, which incorporates
phase transfer, depends
on a good understanding of how the atmospheric path delay changes with
frequency (e.g.\ \citealt{ALMAMemo404}). We
explore how the wet dispersive path delay varies for realistic atmospheric conditions
at the ALMA site using the \atm transmission code. We find the wet dispersive
path delay becomes a significant fraction ($\gtrsim 5$\,per cent) of
the non-dispersive delay for the high-frequency ALMA bands
($\gtrsim 160$\,GHz, Bands 5 to 10). Additionally, the variation
in dispersive path delay across ALMA's 4-GHz contiguous bandwidth is not
significant except in Bands 9 and 10. The ratio of dispersive path delay to
total column of water vapour does not vary significantly
for typical amounts of water vapour, water vapour scale heights and ground pressures above
Chajnantor. However, the temperature profile and particularly the
ground-level temperature are more important. Given the likely constraints from ALMA's ancillary
calibration devices, the uncertainty on the dispersive-path scaling will be around 2\,per cent in the worst case and should
contribute about 1\,per cent overall to the wet path fluctuations at
the highest frequencies.

\end{abstract}

\section{Introduction}

The performance of interferometers at (sub)millimetre wavelengths is
often limited by differential fluctuations in the atmospheric path along
the line of sight to each of the constituent antennas. If uncorrected,
these fluctuations lead to a loss in sensitivity, imaging artefacts
and a limit on the maximum usable baseline
(e.g.\ \citealt{ALMAMemo262,ALMAMemo535}; \citealt*{ALMAMemo582}). ALMA will
correct such path variations using a combination of techniques:
\begin{enumerate}
\item Fast-switching observations of bright calibrator sources
  (e.g.\ quasars).
\item Water vapour radiometry using dedicated 183-GHz radiometers
  (WVRs), installed in every 12-m ALMA antenna. The WVRs will allow the
  retrieval of the amount of water vapour along the line of sight to
  each antenna and thus fluctuations in the atmospheric path length
  resulting from this water vapour.
\item Self calibration, for a fraction of the brightest science targets.
\end{enumerate}
Fast switching interleaves science observations with short ($\sim$ a few seconds) calibration
observations on a cycle time of 20--200\,s. These calibrations allow
an estimate of the atmospheric and instrumental phase
errors \textit{in the direction of the calibrator}. The applicability
of the calibrator phase solutions to the science target depends on
their angular separation and the observing frequency. The radiometric technique will be applied
\textit{continuously}, correcting for atmospheric path fluctuations on timescales shorter
than the fast-switching cycle time, but does require an accurate model
to convert the WVR measurements into path delays. 

Variations in the atmospheric path delay to the antennas predominantly
arise from a combination of fluctuations in the water vapour content \emph{and} density of air
in the troposphere -- respectively, the so-called wet and dry path components. Furthermore, we can
split both components (the wet and dry) into two parts: one dispersive (i.e.\ dependent on frequency)
and the other non-dispersive (independent of
frequency). Conventionally, the non-dispersive part is taken to be the
path in the low-frequency limit. This memo focuses on the dispersive
part of the wet path delay introduced by the atmosphere. Previously,
atmospheric dispersion in the context of fast switching was studied by \citet{ALMAMemo404} and \citet{ALMAMemo523}. They found
the magnitude of the dispersive phase will become non-negligible in ALMA's submillimetre
bands. Furthermore, \citet{ALMAMemo404} quantified the dispersive
phase delay for \emph{typical} conditions at the ALMA site. We extend their work by
investigating how the dispersive path delay depends on the
\emph{variation} of the physical parameters of the atmosphere and what constraints can be placed on it
from the proposed ancillary calibration devices on site. 

This memo is structured into six parts. First, we explain the planned
phase-correction strategy for ALMA and where dispersive effects play a role. Second, in \prettyref{sec:dispersivedelay}, we describe how dispersive and
non-dispersive path delays arise in the atmosphere and how we
compute the dispersive path delay. \prettyref{sec:dispersivedelay} also
quantifies the dispersive path delay across the full range of ALMA
observing frequencies, before \prettyref{sec:physicalinfluences}
demonstrates how varying different atmospheric parameters changes the
magnitude of the dispersive path delay. Finally, we look at what constraints
can be placed on the dispersive path delay with temperature measurements
from weather-monitoring devices.   

\section{Phase calibration for ALMA} \label{sec:almaphasecal}

Before we consider the magnitude of the dispersive path delay, we
will detail the current phase-correction plans for ALMA and how these are affected by atmospheric dispersion.

Numerous previous memos have discussed fast-switching phase correction
for ALMA
(e.g.\ \citealt{ALMAMemo262,ALMAMemo403,ALMAMemo523}). Briefly, fast switching
removes some of the antenna-based phase fluctuations by regularly
making short observations of a bright point-source calibrator, close
to the science target in the sky. This difference in direction between
the target and calibrator means the calibration
solutions need to be interpolated to estimate the science target's
phase errors.

Typically, at ALMA's highest observing frequencies, it will be unlikely that the
calibrations can be taken at the same frequency as the science
observations, for the following reasons:
\begin{enumerate}
\item Calibrators of the necessary strength at the science frequency
  will probably be too far from the science target (although this
  depends on the source counts at high frequencies, which are
  currently not well known).
\item The combination of high frequencies and long baselines may
  resolve many potential calibrators.
\item The phase errors may prove to be so large that phase wrapping
  makes reliable solutions difficult.  
\end{enumerate}   
To overcome these difficulties, one proposal is for ALMA to observe
calibration sources at 90\,GHz when necessary and then scale the corresponding
phase solutions to the science frequency (the \emph{phase-transfer}
technique). The frequency above which the phase transfer-approach
will be necessary depends on the array configuration,
atmospheric conditions and ultimately experience gathered at the site. It is
currently expected to be the routine mode at wavelengths below 2\,mm
(frequencies $\gtrsim 150$\,GHz). In the absence of dispersion, the
required phase scaling is simply the ratio of the calibration and
science frequencies. However, at high frequencies, as we have already
mentioned and will detail below, the numerous nearby water vapour
lines ensure that atmospheric dispersion will need to be taken into account. 

Fluctuations in the atmospheric path on timescales shorter than the
fast-switching cycle time will be corrected radiometrically using WVRs installed in
every 12-m ALMA antenna. The WVRs measure the brightness of the
183-GHz atmospheric water line, which is very sensitive to the total
amount of water vapour in the atmosphere
\citep{ALMAMemo496,ALMAMemo587}. The fundamental difficulty in the analysis of WVR
data is how to convert fluctuations in the measured sky brightnesses
around the 183-GHz water line into
variations in the atmospheric path delay. We have begun to develop a
Bayesian framework for computing such conversion factors
\citep{ALMAMemo587,ALMAMemo588}, which naturally incorporates the prior knowledge of the system and all the observable data. In the first
memo \citep{ALMAMemo587}, we begin with the simplest possible model
atmosphere, comprising a single, thin layer of water, which is the
only cause of path fluctuations. Even so, an analysis of test data from
prototype WVRs that were installed on the Submillimeter Array (SMA) yields corrections that are within $\sim5$\,per cent of optimal ones. Memo 588 \citep{ALMAMemo588} extends this scheme to
include the observed correlation between phase and sky brightness,
which ALMA may implicitly record during fast switching, when the calibrator's phase is measured whilst data are taken with the
WVRs. The inclusion of this empirical relationship should significantly improve the accuracy
of the phase correction. 

Typically, the calculation of phase-correction coefficients (i.e.\
the conversion factors above) from WVR data requires a model of the
atmosphere, which in
principle can be used to estimate the dispersive as well as
non-dispersive path delays. In our
work to date, which was based on relatively low-frequency (220\,GHz)
data, we have not modelled the dispersive path contribution. Instead,
we simply scale the non-dispersive path delay by a constant factor to
estimate the total wet delay. However, in reality this factor is a
function of frequency. Furthermore, if the dispersive
contribution is a significant fraction of the total path delay, it
must be determined with good accuracy. To date there has been little
work on what atmospheric properties influence the magnitude of the
dispersive path delay and by what amount. In
\prettyref{sec:physicalinfluences}, we therefore examine the variation of the
dispersive path delay using multiple realistic models of the
atmosphere above the ALMA site. 

\section{The dispersive path delay} \label{sec:dispersivedelay}

Fluctuations in the wet path
delay, $\rmn{d} l_\rmn{H_2O}$, are separated into the sum of non-dispersive ($\rmn{d}
s$) and dispersive delays ($\rmn{d}S_\nu$):

\begin{equation} \rmn{d}l_\rmn{H_2O} = \rmn{d}s + \rmn{d}S_\nu
  . \label{eqn:sumofparts}\end{equation} The \emph{total} non-dispersive path delay (i.e.\ the sum of the wet
and dry components) can be computed from the Smith-Weintraub equation for the refractive
index, $n$, at a temperature, $T$ \citep{ALMAMemo517,ALMAMemo587}:
\begin{equation}
n-1 = 10^{-6} \left[ \alpha \frac{P_\rmn{d}}{T} + \beta
  \frac{P_\rmn{w}}{T} + \gamma \frac{P_\rmn{w}}{T^2}\right]\rmn{,} 
\end{equation} 
where $P_\rmn{d}$ and $P_\rmn{w}$ are the partial pressures of the dry
air and water vapour respectively and $\alpha$, $\beta$ and $\gamma$
are constants. Since we are only concerned with the excess path
introduced by water vapour (i.e.\ d$s$ above), then we may omit the
first term and of the remaining terms the last one dominates. Hence we
can transform to the following expression \citep{ALMAMemo587}:\begin{equation} \rmn{d}s \approx
  \frac{1741\,\rmn{K}}{T}\rmn{d}c
  , \label{eqn:nondispersivedelay} \end{equation} where $c$ is the
water vapour column.

In general, atmospheres with spectral lines are dispersive,
with the dispersion related to the absorption according to the
Kramers-Kr\"onig relations. Thus, any atmospheric property that
affects the shapes of absorption-line features will probably influence
the atmospheric dispersion, summarized in the following functional form (see
\prettyref{tab:atmparameters} for a list of the symbols used): 

\begin{equation} \rmn{d}l_\rmn{H_2O} \approx \left
    (\frac{1741\,\rmn{K}}{T}\right ) \rmn{d}c + \rmn{d}S_\nu(\nu,T,\Gamma,p,c) , \label{eqn:totaldelay}\end{equation}
where we have substituted the non-dispersive delay from
\prettyref{eqn:nondispersivedelay} into \prettyref{eqn:sumofparts}. The variation of d$S_\nu$ as a
function of its parameters is studied in the remainder of this memo. 

\subsection{Computing the dispersive path delay}

We calculate the dispersive and non-dispersive contributions to the wet and dry path
delays between 1\,GHz and 1\,THz from model atmospheres using the
\atm software \citep*{pardo01}\footnote{We have packaged {\sc atm}
  as {\sc aatm}, which is available under the GPL license from
  \url{http://www.mrao.cam.ac.uk/~bn204/alma/atmomodel.html}. We
  interface with the \atm libraries using the \texttt{dispersive}
  program \citep{ALMAMemo587}, from version 0.1 of \aatm.}. \atm
accurately predicts the atmospheric opacity above Mauna Kea, Hawai'i
up to 1.6\,THz \citep{pardo05}. Its dispersive calculations have not
been as thoroughly verified due to a lack of high-frequency
test data, which should change in the near future once the ALMA WVRs
become operational at Chajnantor. 

\begin{figure*}
\includegraphics[width=14cm]{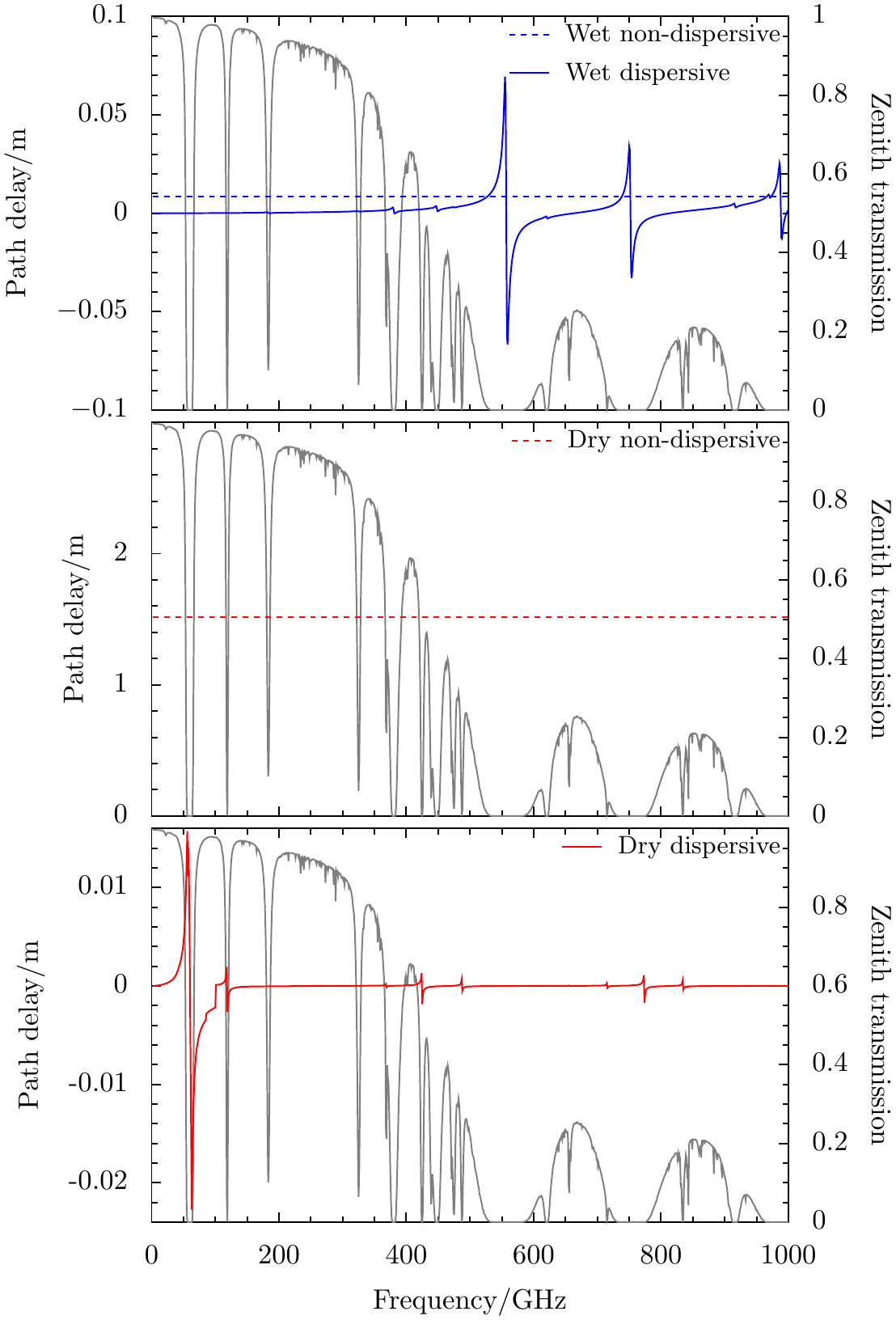}
\caption{Wet non-dispersive (\textit{top panel, dashed blue}) and dispersive
  (\textit{top panel, solid blue}) path delays alongside the dry
  non-dispersive (\textit{middle panel, dashed red})
  and dispersive (\textit{bottom panel, solid red}) path delays calculated by \atm for the AOS
  model atmosphere with standard parameters as listed in
  \prettyref{tab:atmparameters}, including a 1.22\,mm column of
  water. Overlaid in grey on all three panels is the corresponding
  zenith atmospheric transmission, calculated from the total opacity,
  $\tau$ (as $e^{-\tau}$).}
\label{fig:baseconditions}
\end{figure*}

In \prettyref{fig:baseconditions}, we plot the wet and dry, dispersive
and non-dispersive path delays overlaid on top of the atmospheric
transmission, all computed using the basic model parameters
listed in \prettyref{tab:atmparameters}, which are suitable for the ALMA array
operations site (AOS) at Chajnantor. The parameters listed in
\prettyref{tab:atmparameters} are the median values measured by
site-monitoring equipment where possible and their
origin will be detailed in later sections. The computed dispersive path delay is small except around the atmospheric \htwoo lines
(for the wet component) or \otwo lines (for the dry). The largest contribution to the total delay at all frequencies
arises from the dry non-dispersive path delay. However, we expect the dry
column to be relatively stable and so contribute very little to the
\textit{differential} path delay \citep{ALMAMemo404,ALMAMemo523} i.e.\
the difference in delay between antennas. We will ignore the dry path delays for the rest of
this memo but they may prove to be a significant extra source of error
that remain after phase correction using the WVRs. 

\begin{table}
\caption{Basic parameters of the atmospheric model of the ALMA AOS
  site, computed from median site characterization data except $p_\rmn{g}$
  (see \prettyref{apdx:sitedata} for details).}
\label{tab:atmparameters}
\begin{tabular}{cccc}
\toprule
Parameter & Units & Value & Comment\\
\midrule
$c$        & mm   & 1.22   & Zenith water column\\
$p_\rmn{g}$ & mb & 560 & Ground-level pressure\\
$T_\rmn{g}$ & K    & 270 & Ground-level temperature\\
$\Gamma_\rmn{T}$   & K\,km$^{-1}$ & $-7.28$ & Tropospheric lapse rate\\
$h_0$      & km   & 1.16   & Water vapour scale height\\    
\bottomrule
\end{tabular}
\end{table}

\begin{figure*}
\includegraphics[width=\textwidth]{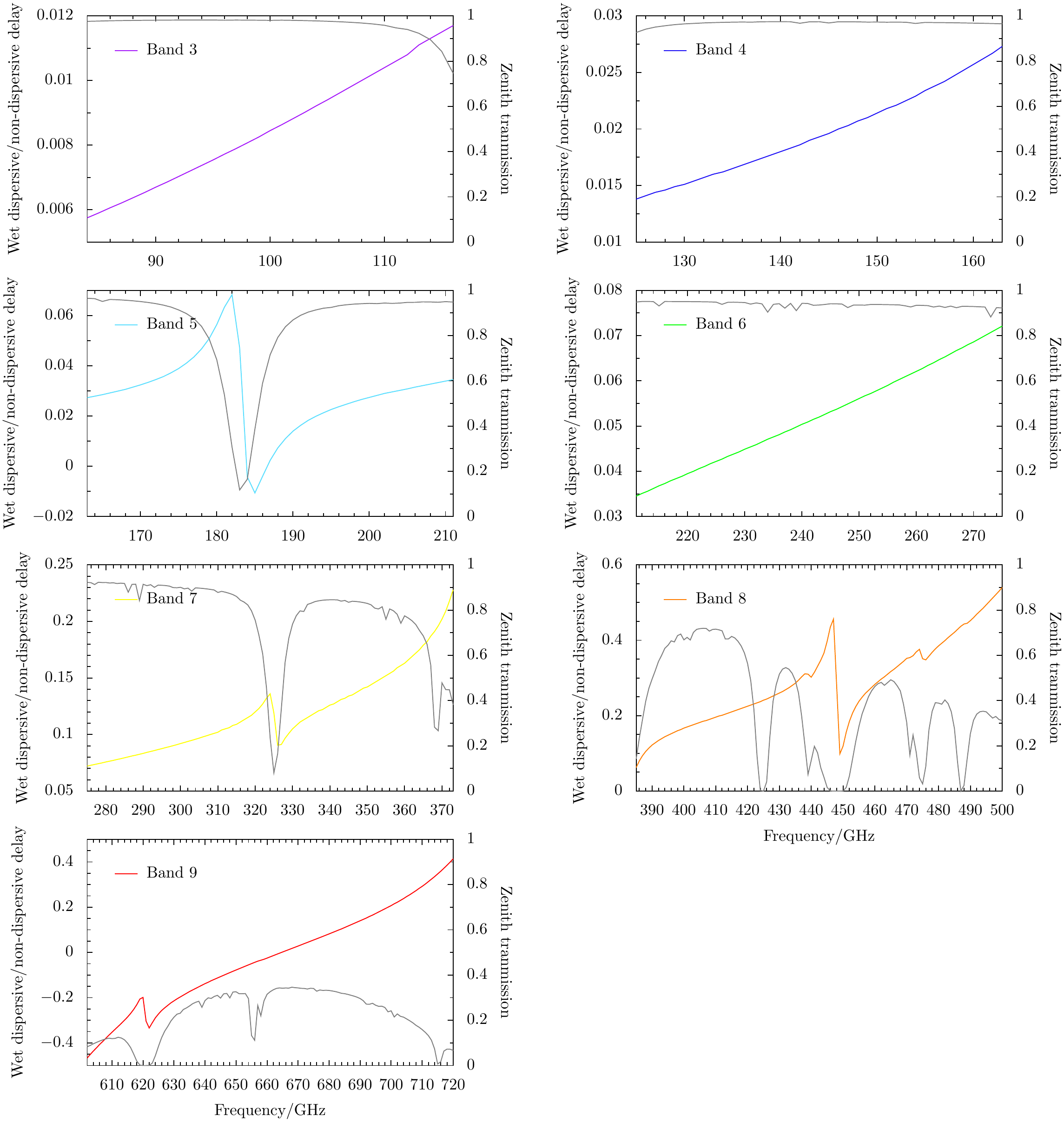}
\caption{Ratio of the wet dispersive to wet non-dispersive delay for
  the construction ALMA observing bands, listed in \prettyref{tab:almabands}. Also plotted (\textit{grey}) is the
  atmospheric zenith transmission at the same frequency. The ratio
  rises continually with frequency and becomes particularly
  significant in the submillimetre bands (Bands 7, 8 \& 9).}
\label{fig:phaseratio}
\end{figure*}

In \prettyref{fig:phaseratio}, we plot the ratio of wet dispersive to
non-dispersive path delay across the (currently-funded) ALMA
observing bands (see \prettyref{tab:almabands}). In the first
two of these bands (Bands 3 and 4), the dispersive path delay is between 0.5 and 3\,per
cent of the non-dispersive, rising as a fraction approximately
linearly with frequency. At the frequencies of Band 5 and higher, the dispersive path
becomes a more significant fraction of the total wet delay ($\gtrsim 5$\,per cent)
and will need to be considered in both the fast-switching and WVR
analyses. In Band 8 for instance, as far as possible from the
absorption lines, the dispersive path delay is 20 to 40\,per cent of the
non-dispersive. 

\begin{table}
\caption{ALMA receiver bands. In the initial phase of operations ALMA
  will be equipped with only six bands: 3,4, \& 6--9. Bands 1 and 2 are not
yet funded.}
\label{tab:almabands}
\begin{tabular}{l l l l l l}
\toprule
Band  & Frequency   & $T_\rmn{rx}$~$^1$ & $\nu_\mathrm{rep}$~$^2$ & Mixing~$^3$ & Supplier~$^4$ \\
      & range (GHz) & (K) & (GHz) & Scheme \\                      
\midrule
   1  & 31--45   & 17  & 38  & USB & --\\ 
   2  & 67--90   & 30  & 79  & LSB & --\\
   3  & 84--116  & 37  & 100 & 2SB & HIA \\   
   4  & 125--163 & 51  & 144 & 2SB & NAOJ \\
   5  & 163--211 & 65  & 200 & 2SB & OSO$^\dag$ \\
   6  & 211--275 & 83  & 243 & 2SB & NRAO \\
   7  & 275--373 & 147 & 342 & 2SB & IRAM \\
   8  & 385--500 & 196 & 405 & 2SB & NAOJ \\
   9  & 602--720 & 175 & 680 & DSB & NOVA \\
   10 & 787--950 & 230 & 869 & DSB & NAOJ \\
\bottomrule
\end{tabular}\\
\begin{small}
$^{1}$~Receiver noise temperature specification for over 80\,per cent
of the band.\\
$^2$~Representative frequency, either the band centre or where
  the transmission is better if the centre is near an absorption
  line.\\
$^3$~Two lowest-frequency bands use HEMT mixer technology and are single sideband,
either upper (USB) or lower (LSB); all the others use SIS mixers and are
either dual sideband (2SB -- each sideband detected separately) or
double sideband (DSB).\\
$^4$~HIA -- Herzberg Institute of Astrophysics, Canada; NAOJ --
National Astronomical Observatory of Japan; OSO -- Onsala Space
Observatory/ Chalmers University, Sweden; NRAO -- National Radio
Astronomy Observatory, USA; IRAM - Institut de Radio Astronomie
Millim\'etrique, France; NOVA -- Nederlandse Onderzoekschool voor de Astronomie.\\
$^\dag$~Six production receivers will be provided through the European
Commission's Framework 6 programme.   
\end{small}                                    
\end{table}

\begin{figure}
\includegraphics[angle=90,width=8cm]{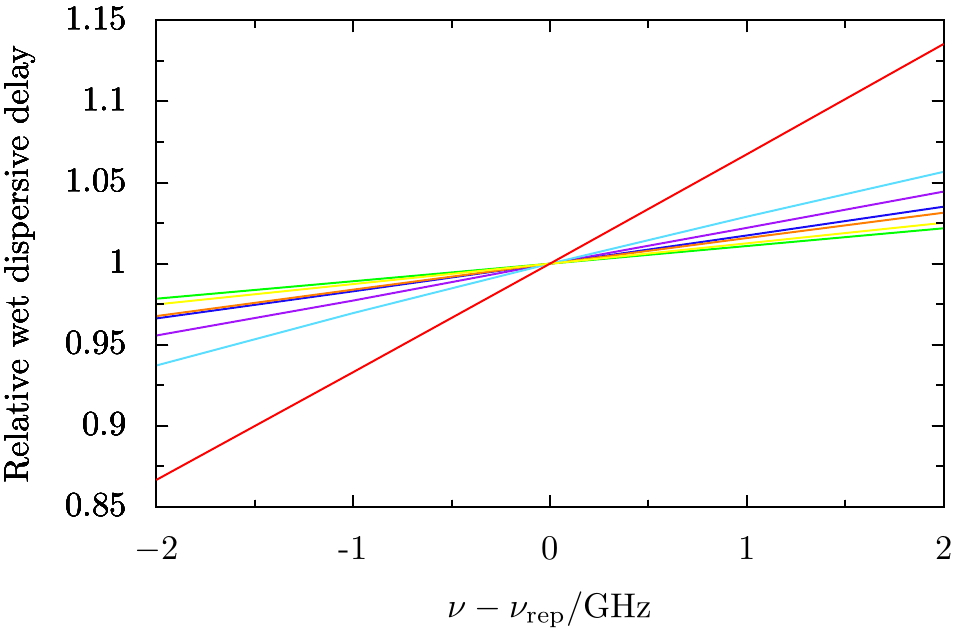}
\caption{Variation in wet dispersive path delay for the construction
  ALMA bands, which were listed in \prettyref{tab:almabands}. The lines use the
  same colours as \prettyref{fig:phaseratio} for the different bands: 3
  (\textit{purple}), 4 (\textit{dark blue}), 5 (\textit{light blue}),
  6 (\textit{green}), 7 (\textit{yellow}), 8 (\textit{orange}), 9
  (\textit{red}). The delays are plotted for frequencies in a 4-GHz
  portion around the bands' representative frequencies (listed in
  \prettyref{tab:almabands}, typically the band centres). This portion
  represents half of the instantaneous contiguous bandwidth (either
  the upper or lower sideband) for the ALMA 2SB mixers. Each measurement is divided by the value at the
  representative frequency and all the computations are for the median atmosphere
 specified in \prettyref{tab:atmparameters}. The Band 9 receivers are
DSB and therefore provide 8-GHz contiguous bandwidth (not
plotted), which results in about a 30\,per cent variation in the
dispersive path delay towards the band edges compared with the representative frequency.}
\label{fig:almabandwidth}
\end{figure}

Additionally, we can investigate the phase slope that the atmosphere would
introduce across the observing bandpass if no account were taken for
variations in the dispersive path delay with frequency. In
\prettyref{fig:almabandwidth}, we plot the fractional variation in the
dispersive path delay for a 4-GHz portion of frequency space around representative
band frequencies listed in \prettyref{tab:almabands}. For each band we
have normalized the values with respect to the chosen bandpass centre frequency. As
we noted for \prettyref{fig:phaseratio}, the dispersive path delay varies approximately linearly for small frequency widths. For most of the
bands the variation is $\pm 2-5$\,per cent over the observing
bandwidth. Thus, such variation can be ignored, particularly at the lowest
frequencies where the dispersive path delay makes a small contribution
overall. Of the bands plotted, only
Band 9 exhibits large variations of some $\pm 13$\,per cent across the
4-GHz chunk, which
rises to nearly $\pm 30$\,per cent if the full 8-GHz DSB bandwidth is
considered. In this case, a frequency-dependent scaling or multiple independent
calculations of the dispersive phase (as planned in the ALMA
correlator and TelCAL software sub-system) will be important. 

\section{Physical influences on the dispersive path delay} \label{sec:physicalinfluences}

In this section we quantify the impact of changes in atmospheric
conditions on the wet dispersive path delay. We
explore the influence of: the quantity of water vapour (\prettyref{sec:varyingc}), the
ground-level temperature (\prettyref{sec:varyingT}) and pressure
(\prettyref{sec:varyingp}) alongside the distribution of temperature
with height (\prettyref{sec:varyingtlr}) and the
distribution of water vapour with height (\prettyref{sec:varyingh0}). For each investigation we
compute the dispersive delay from \atm using a variety of atmospheric
models designed to represent average and extreme conditions at the
ALMA AOS. 

Our focus is the radiometric phase-correction technique using the
WVRs. An important requirement for this technique is the optimal
computation of time-dependent phase-correction coefficients,
d$L/$d$T_{\rmn{B},i}$, for each of the four WVR channels. These coefficients relate
a change in atmospheric path, $\delta L$, to a change in the observed sky brightness by
a WVR, in each of its four channels, $\delta T_{\rmn{B},i}$, via (see
\citealt{ALMAMemo587,ALMAMemo588}):
\begin{equation}\label{eq:phasecalcoefficients} \delta L = \sum_{i=1}^{4} w_i
  \frac{\rmn{d}L}{\rmn{d}T_{\rmn{B},i}} \delta
  T_{\rmn{B},i}\rmn{,}\end{equation} where $w_i$ is an appropriate weighting for
each channel. We can decompose this expression further, since the WVRs
are really only sensitive to variations in
the water vapour column, $\delta c$, which we can relate to the
measured sky brightness by introducing coefficients, d$c/$d$T_{\rmn{B},i}$:
\begin{equation}\label{eq:watercoefficients} \delta c = \sum_{i=1}^{4} w_i
  \frac{\rmn{d}c}{\rmn{d}T_{\rmn{B},i}} \delta
  T_{\rmn{B},i}\rmn{.}\end{equation}
Although the WVRs do place constraints on the atmospheric temperature
and pressure, normally they cannot detect small fluctuations in their
values. We can then write the variation in wet path delay, $\delta l_\rmn{H_2O}$,
as the product of the fluctuation in water vapour content and the sum
of two scaling terms:
\begin{equation} \delta l_\rmn{H_2O} = \left (
    \frac{1741\,\rmn{K}}{T} + \frac{\rmn{d}S_\nu}{\rmn{d}c} \right )
  \delta c.\end{equation}
The first scaling term is the non-dispersive path delay from
\prettyref{eqn:nondispersivedelay}, while the second quantifies the
dispersive path delay. Finally, as we saw in
\prettyref{sec:dispersivedelay}, although d$S_\nu$/d$c$ may vary a lot, it
is only related to the \emph{dispersive path delay}, which in the
millimetre bands makes a very small contribution to the total path delay
compared to the non-dispersive path.

\subsection{Water vapour quantity} \label{sec:varyingc}

\begin{figure*}
\includegraphics[width=0.95\textwidth]{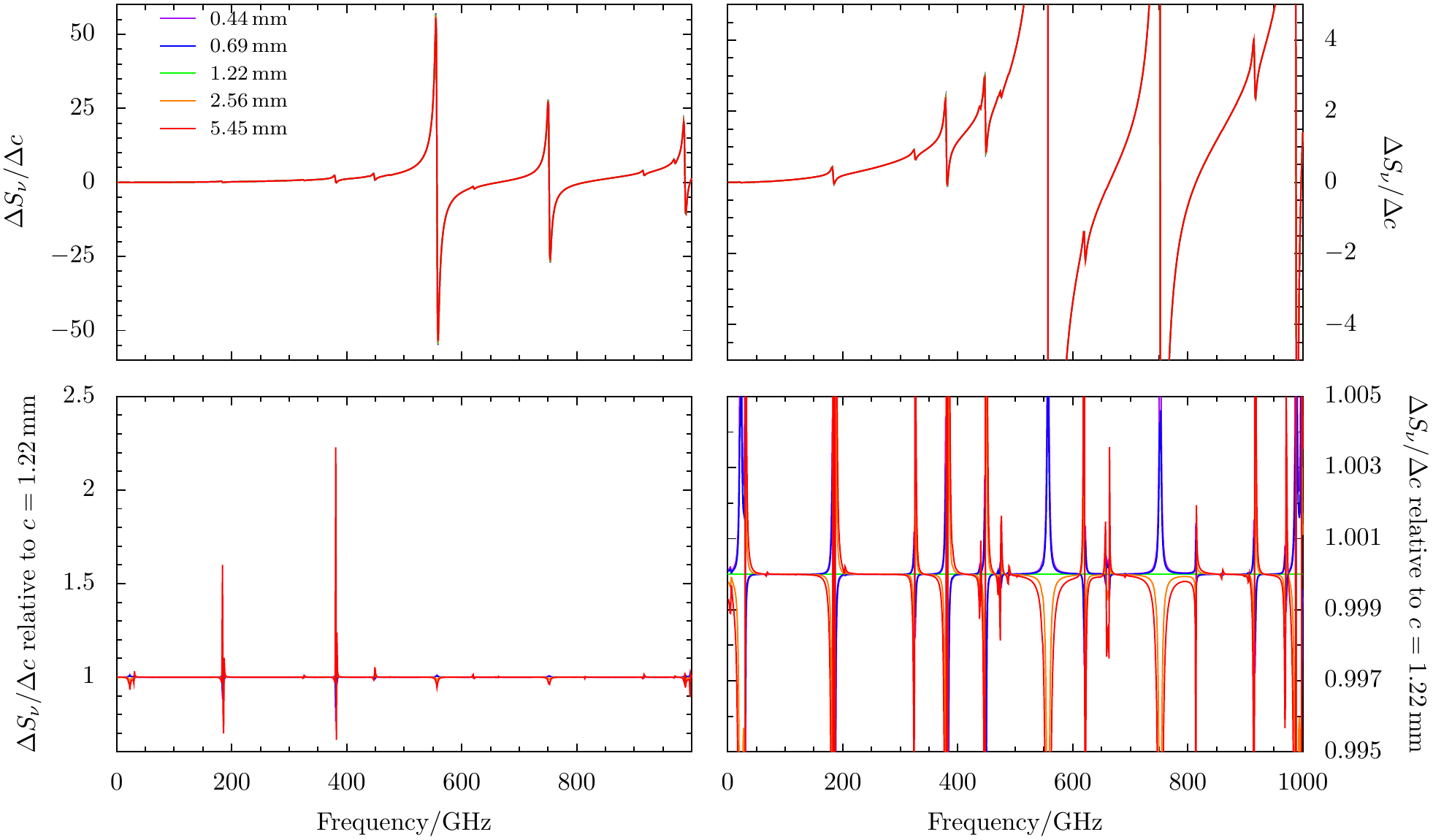}
\caption{Variation of \param with column of water,
  $c$. The upper panels show the absolute variation in \param as a function of frequency from 1 to 1000\,GHz, while the lower panels show the
  variation relative to (i.e.\ divided by) the median profile with $c=1.22$\,mm. The right-hand panels show the same data as in
  left-hand ones but over a narrower range of values. The plotted data
  come from profiles containing 0.44, 0.69, 1.22, 2.56 and 5.45\,mm of water respectively corresponding to
the 10, 25, 50, 75 and 90 percentiles of the PWV cumulative function
at Chajnantor (see \prettyref{sec:watervapourcolumn}). All the curves are plotted in each
of the panels: if only one line is discernible then the curves overlap.}
\label{fig:varyingc}
\end{figure*}

First, we investigate the effect of varying the column of water vapour, $c$, in the standard AOS
atmosphere (\prettyref{tab:atmparameters}). In \prettyref{fig:varyingc}
we plot \param for $c=0.44$, 0.69, 1.22, 2.56 and 5.45\,mm corresponding to
the 10, 25, 50, 75 and 90 percentiles of the precipitable water vapour
(PWV) cumulative function at Chajnantor respectively. These
percentiles are derived from the cumulative 225\,GHz opacity
distributions as described in \prettyref{sec:watervapourcolumn} of
\prettyref{apdx:sitedata}. \prettyref{fig:varyingc} is the template for all
the plots that follow in this section. The upper panels display the
absolute variation in \param for the full extent of the
ordinate (\emph{left panel}) and for an enlarged portion close to the
horizontal axis (\emph{right panel}). In the lower panels, we plot the relative
change in \param, computed by dividing its spectrum
(i.e.\ \param as a function of frequency)
by the one from the median column, $c=1.22$\,mm. Again, on the
left-hand side we show the full ordinate range, whilst on the right we
enlarge the plot around a ratio of unity.  

The shape of the \param spectrum closely resembles that
of the dispersive path delay overall, i.e.\ $\Delta S_\nu$ is
proportional to $\Delta c$ (see \prettyref{fig:baseconditions}),
having discontinuities where the dispersive path delay wraps around, near
the dips of the atmospheric transmission and bright water
lines. Even in the enlarged (\textit{upper right panel}) and relative
plots (\textit{lower panels}) there are
no discernible differences between \param for the different values of
$c$. The only significant departures from unity for the
relative \param are at discontinuities in \param. Correspondingly,
over the entire range of conditions likely at the ALMA
AOS, the amount of water vapour does not affect \param, only supplying
a non-dispersive linear change. Furthermore, varying the
  telescope's elevation/line of sight airmass has an equivalent effect to altering the water
  vapour column at zenith. Therefore our plots also indicate that the
  dispersive phase scaling does not depend on airmass and will not
  need to be recomputed for changes in elevation during observing. 

\subsection{Air temperature at ground level} \label{sec:varyingT}

\begin{figure*}
\includegraphics[width=0.95\textwidth]{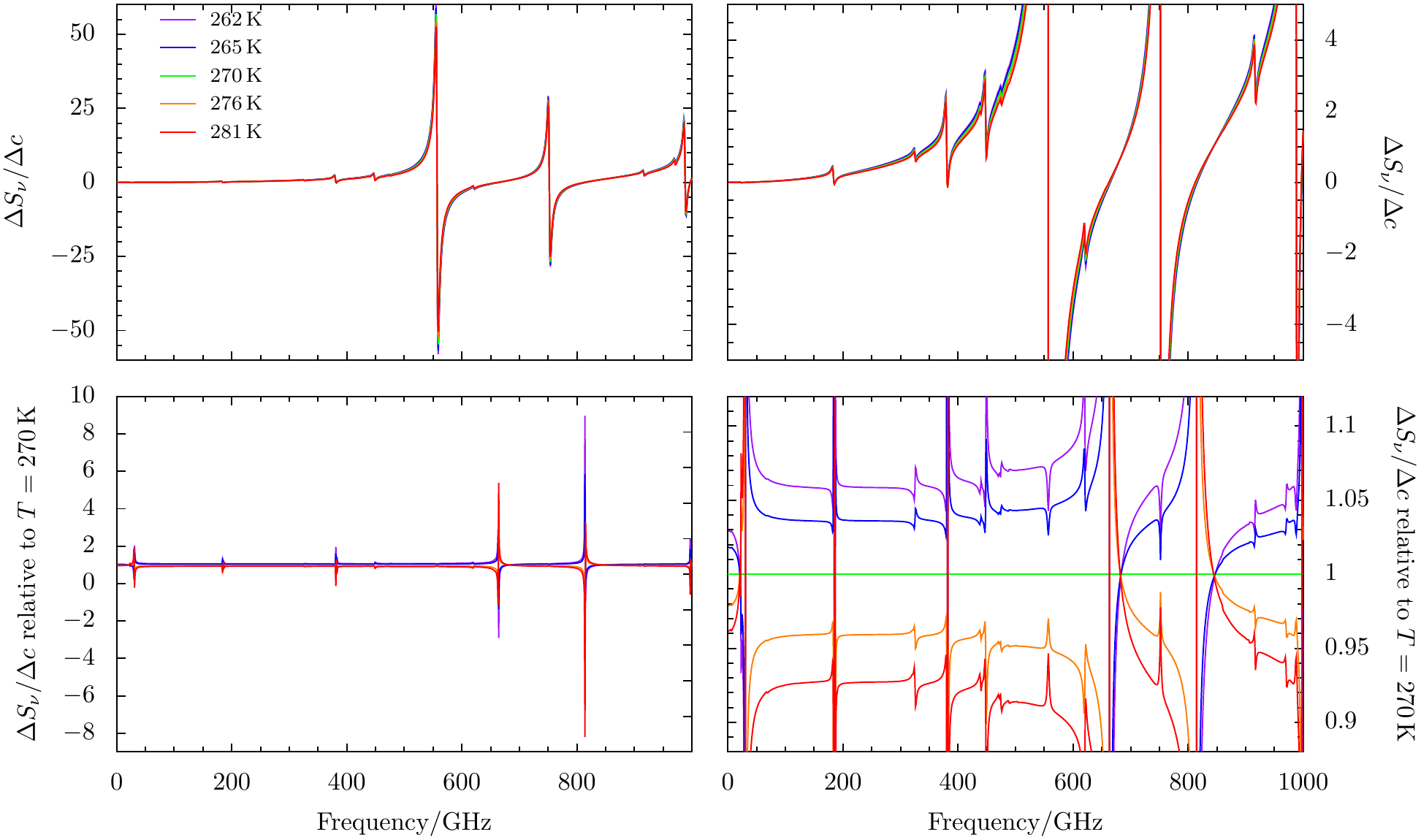}
\caption{Variation of $\Delta L/\Delta c$ with ground temperature,
  $T_\rmn{g}$. Panels are laid out as for \prettyref{fig:varyingc}. The
  plotted data have $T_\rmn{g}$ of 262, 265, 270, 276, 281\,K corresponding to
the 10, 25, 50, 75 and 90 percentiles of the temperature cumulative function
at Chajnantor. The temperature data are from \citet{sitecharacterization}.}
\label{fig:varyingT}
\end{figure*}

Next, we look at how the ground-level temperature affects the predicted
phase-correction coefficients. As in the previous section
(\prettyref{sec:varyingc}), we varied $T_\rmn{g}$, in the standard AOS
atmosphere which was used as the input to \atm. \prettyref{fig:varyingT} provides our results for \param, in the same format as \prettyref{fig:varyingc}, using
$T_\rmn{g}=262$, 265, 270, 276 and 281\,K corresponding to the 10, 25,
50, 75 and 90 percentiles of the temperature cumulative function at
Chajnantor \citep{sitecharacterization}. 

The key plot is in the lower right panel, where we plot the spectrum
for each $T_\rmn{g}$ relative to 270\,K. The relative \param computed
are constant with frequency except at the centres of absorption
features, i.e.\ the same frequencies as in
\prettyref{fig:varyingc}. The variation in the ratio at
different \tg is higher than for the different water vapour columns,
and therefore will have more significance. Between the 25 and 75 percentiles, the difference in \param is
5--6\,per cent but at the 10 and 90 percentiles it becomes
7--9\,per cent. These differences are likely to be non-negligible for
phase correction at submillimetre frequencies.

This variation is also important if we recall the strong diurnal
change in temperature at Chajnantor. \citet{ALMAMemo541} found the
temperature varied between approximately $-5$ and $5^\circ$C at their
site A over 24 hours. Such a change could alter the dispersive path
delay scaling by around $\pm 5$\,per cent. We return to the
constraints we can place on this source of uncertainty using meteorological data from the ALMA
ancillary calibration devices in \prettyref{sec:constraints}.

\subsection{Air pressure at ground level} \label{sec:varyingp}

\begin{figure*}
\includegraphics[width=0.95\textwidth]{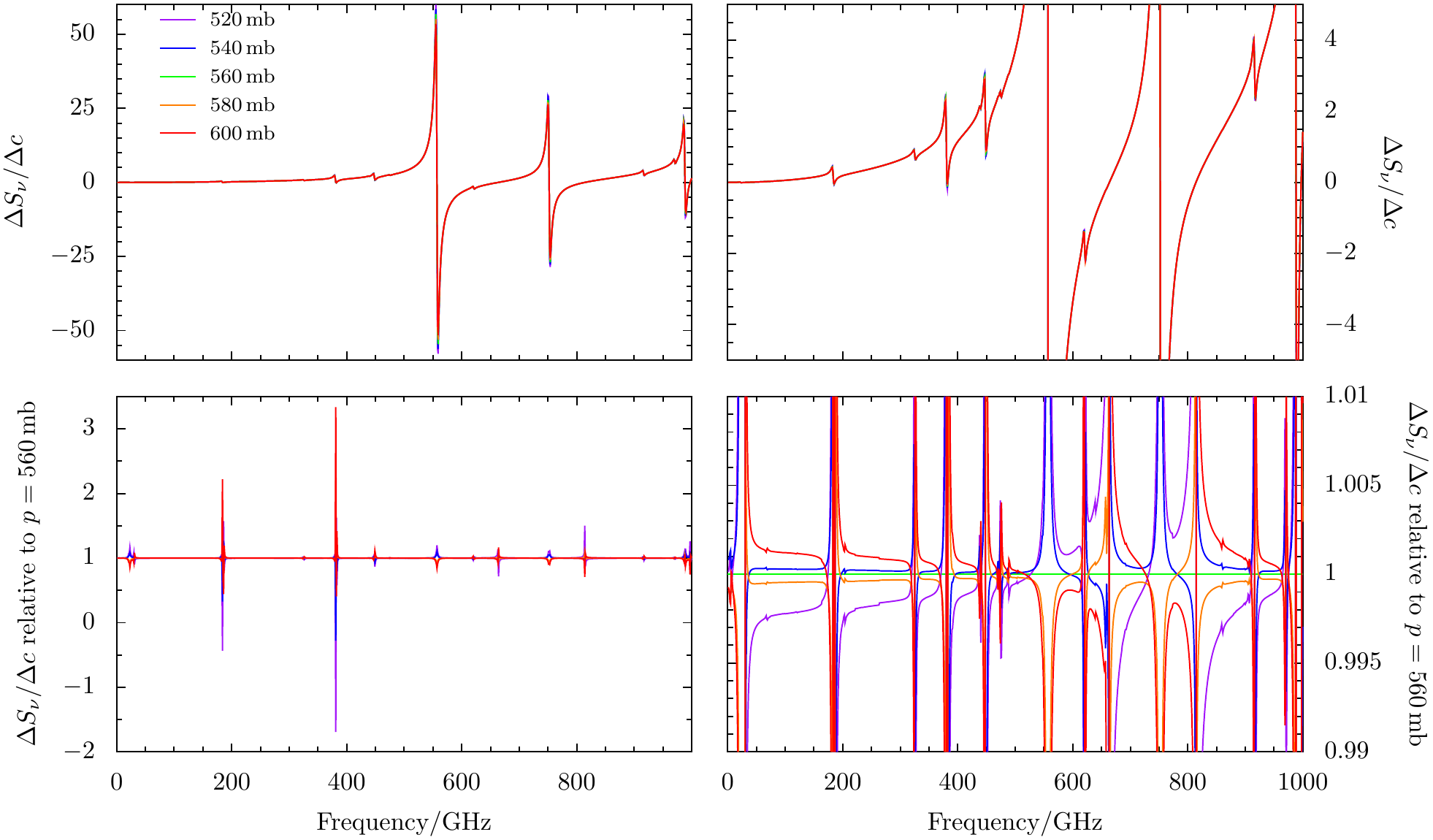}
\caption{Variation of $\Delta L/\Delta c$ with ground pressure,
  $p_\rmn{g}$. Panels are laid out as for \prettyref{fig:varyingc}. The
  plotted data have $p_\rmn{g}$ of 520, 540, 560, 580, 600\,mb.}
\label{fig:varyingp}
\end{figure*}

We have checked the impact of changing the air pressure at ground
level, $p_\rmn{g}$, by repeating the previous \atm calculations using our
standard AOS atmosphere with $p_\rmn{g}$ set to 520, 540, 560, 580,
600\,mb. The results, shown in \prettyref{fig:varyingp}, indicate that
\param varies $p_\rmn{g}$ at a level typically less than 0.2\,per
cent. This is much smaller than the effect resulting from any of the other atmospheric
parameters we consider, so $p_\rmn{g}$ is not a likely source of
uncertainty in the WVR phase-correction coefficients.
 
\subsection{Tropospheric lapse rate} \label{sec:varyingtlr}

\begin{figure*}
\includegraphics[width=0.95\textwidth]{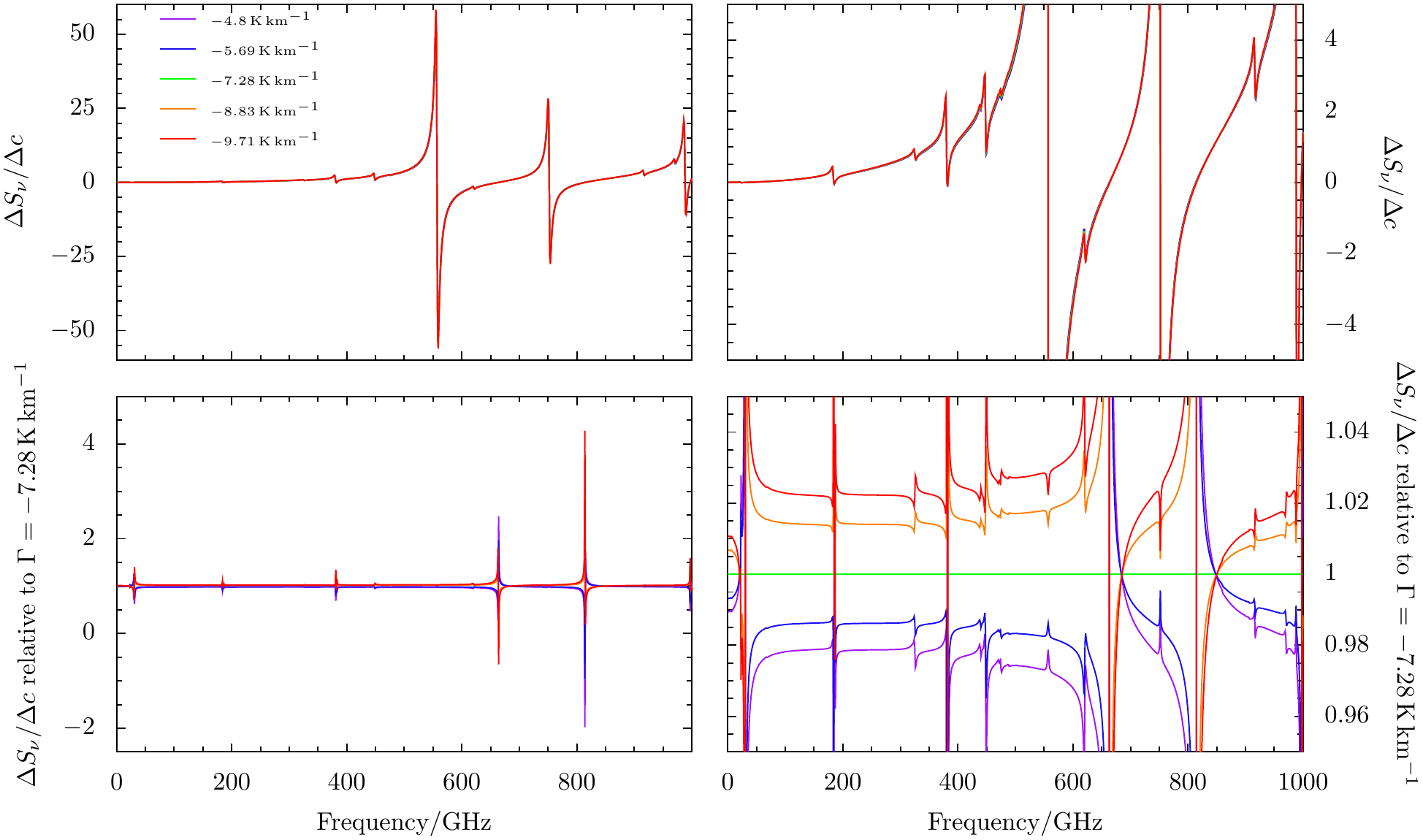}
\caption{Variation of $\Delta L/\Delta c$ with tropospheric lapse rate, $\Gamma_\rmn{T}$. Panels are laid out as for \prettyref{fig:varyingc}. The
  plotted data have $\Gamma_\rmn{T}$ of $-4.80$, $-5.69$, $-7.28$, $-8.83$,
  $-9.71$\,K\,km$^{-1}$ corresponding to
the 10, 25, 50, 75 and 90 percentiles of the \tlr cumulative function
at Chajnantor (see \prettyref{sec:verticalparams}). $-10.0$\,K\,km$^{-1}$ is the dry adiabatic
  lapse rate found in very dry atmospheres.}
\label{fig:varyingtlr}
\end{figure*}

The main way in which we parameterize the vertical temperature
distribution of the model atmospheres is through the tropospheric lapse rate,
$\Gamma_\rmn{T}$. We investigate the effect of variations in \tlr on the dispersive path delay in \prettyref{fig:varyingtlr}, 
where \param is plotted for $\Gamma_\rmn{T}=-4.80, -5.69, -7.28,
-8.83, -9.71$\,K\,km$^{-1}$.  These $\Gamma_\rmn{T}$ again
correspond to the 10, 25, 50, 75 and 90 percentiles of the \tlr
distribution, which we derive from radiosonde data as explained in
\prettyref{sec:verticalparams}. Realistic variations in the lapse rate
at the AOS will produce small, non-negligible changes in the dispersive path scaling of some
2--3\,per cent from median values.  Later in \prettyref{sec:constraints}, we
look at how well we can constrain the lapse rate and thus the path
scaling using the proposed atmospheric temperature profiler for ALMA.

\subsection{Scale height of atmospheric water vapour} \label{sec:varyingh0}

\begin{figure*}
\includegraphics[width=0.95\textwidth]{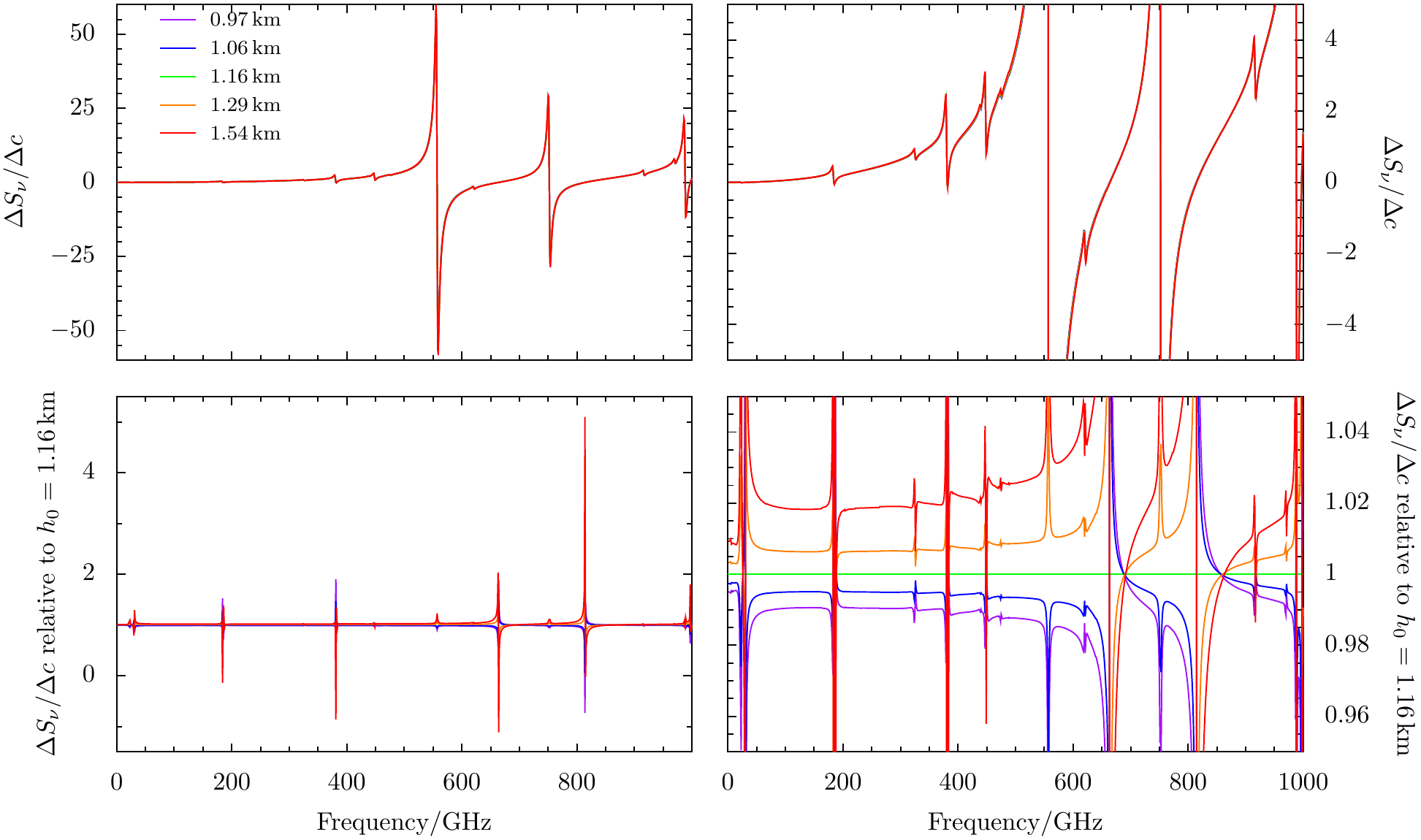}
\caption{Variation of $\Delta L/\Delta c$ with water vapour scale
  height, $h_\rmn{0}$. Panels are laid out as for
  \prettyref{fig:varyingc}. The
  plotted data have $h_0$ of 0.97, 1.06, 1.16, 1.29, 1.54\,km corresponding to
the 10, 25, 50, 75 and 90 percentiles of the water vapour scale height cumulative function
at Chajnantor. }
\label{fig:varyingh0}
\end{figure*}

The final parameter we interogate is the water vapour scale height,
$h_0$. \param is plotted in \prettyref{fig:varyingh0} for $h_0=0.97$,
1.06, 1.16, 1.29, 1.54\,km corresponding to the 10, 25, 50, 75 and 90
percentiles of the water vapour scale height cumulative function at
Chajnantor (see \prettyref{sec:verticalparams}). The relative
dependence of \param on $h_0$ is again similar to the other
parameters, i.e.\ it has spikes at the centres of absorption features. The
relative \param typically varies from the median spectrum by between
0.5 and 1\,per cent, which rises to 2--4\,per cent in the 10 and 90
percentile cases. Thus, under the typical range of conditions at the site,
$h_0$ is a small source of uncertainty in the dispersive scaling for phase correction.

\section{Constraints from the ancillary calibration devices} \label{sec:constraints}

Our predictions from the atmospheric models above indicate that the natural
variation of certain atmospheric properties at the ALMA site could
introduce non-negligible errors in the estimation of phase-correction
coefficients and also in the phase-transfer scaling. However, ALMA
will be equipped with a suite of ancillary calibration devices, including five meteorological
towers and a temperature profiler, which will monitor basic
atmospheric parameters such as the pressure, temperature and wind
speed. In this section, we look at what constraints the temperature
devices should be able to place on the dispersive path delay
prediction.

First, we examine the ground temperature measurement. Although
temperature probes will be positioned on every meteorological tower
and possibly on some of the antennas, we are unlikely to be able to
measure $T_\rmn{g}$ very accurately for
each antenna. This is mainly because the temperature profile for the first 100\,m of the
atmosphere above the ALMA site is strongly controlled by surface heating
and cooling, which results in variations of $\pm 5$\,K over the
site \citep{ALMAMemo541}. Therefore, we estimate we will be able to
measure \tg to around $\pm 2$\,K for each antenna. 

\begin{figure}
\includegraphics[width=8cm]{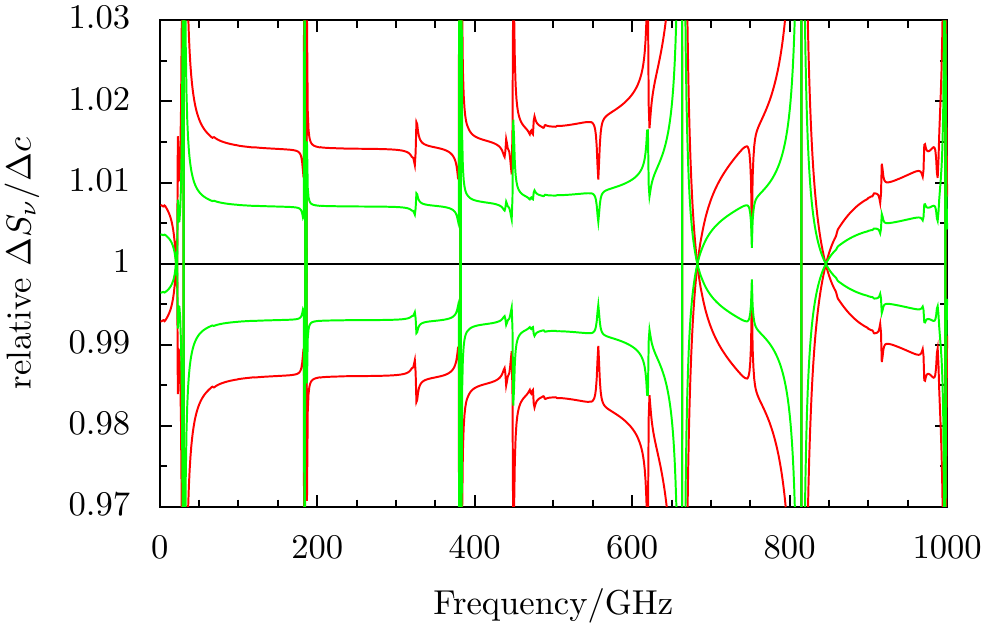}
\caption{Constraints placed on the dispersive \param from a
  measurement of $T_\rmn{g}$. The lines plot the relative change
  in \param from the median \tg (270\,K) spectrum, when \tg is changed
  by 1\,K (\emph{green}) or 2\,K (\emph{red}).}
\label{fig:groundTconstraints}
\end{figure}

Using the standard AOS atmosphere with $T_\rmn{g}=270$\,K
as the input to \atm, we computed the relative change in \param for a
1 and 2\,K change in $T_\rmn{g}$, which we plot in
\prettyref{fig:groundTconstraints}. At frequencies where the
dispersive phase makes a significant contribution to the overall path
delay ($\gtrsim 345$\,GHz), an estimate of the ground temperature to
$\pm 2$\,K can constrain the dispersive phase-correction
coefficient to $\pm 1.5-2.0$\,per cent. This is just under the limit of what would
be acceptable in a total phase calibration budget of 2\,per cent. If
we could do better and constrain \tg to $\pm 1$\,K, then we could get
to better than a 1\,per cent error. 

\begin{figure}
\includegraphics[width=8cm]{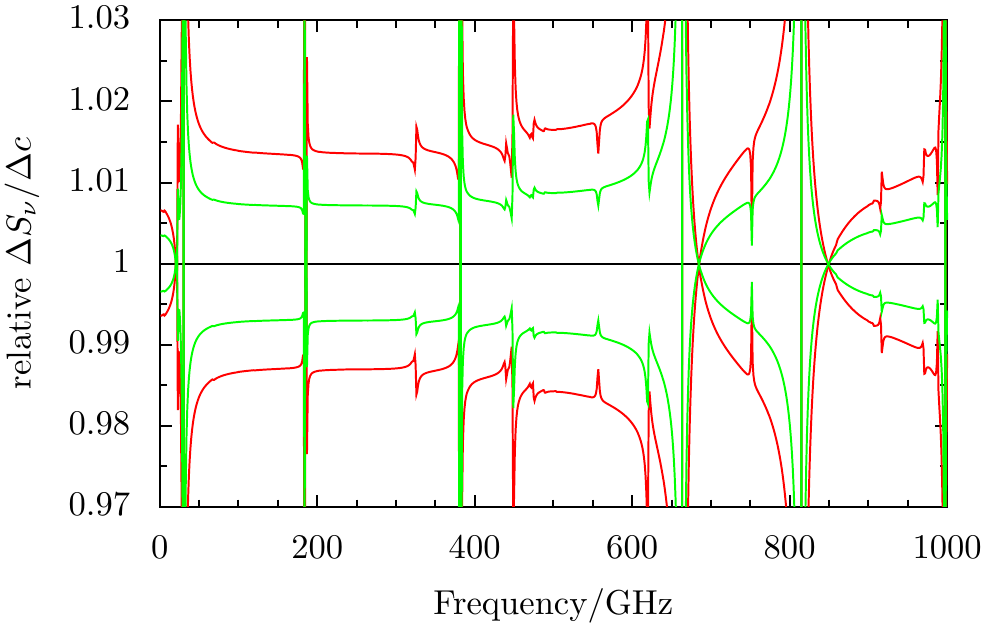}
\caption{Constraints placed on the dispersive \param from a
  measurement of lapse rate by the atmospheric temperature
  profiler. The lines show the relative change
  in \param from the typical $\Gamma_\rmn{T}= -7.28$\,K\,km$^{-1}$ spectrum, when $\Gamma_\rmn{T}$ is changed by
0.8\,K\,km$^{-1}$ (\emph{green}) or 1.5\,K\,km$^{-1}$ (\emph{red}).}
\label{fig:profilerconstraints}
\end{figure}

ALMA also plans to utilize an atmospheric temperature profiler, which
will use multi-frequency observations of the sky brightness
around the 60\,GHz \otwo lines to infer the atmospheric temperature
profile. Such a profile will provide information about the air
temperature away from the ground where surface effects are
minimized and the values are probably more settled. The current
specifications of the profiler unit state it will measure the
atmospheric temperature to $\Delta T\leq 2$\,K with a vertical resolution of $\Delta z \leq
200$\,m up to 1500\,m above the AOS. We assume the temperature profile
consists of 8 measurements of the temperature separated vertically by
200\,m, and each accurate to $\pm 2$\,K. If the profile
is a straight line, we can perform a least-squares fit, which should measure the lapse rate, $\Gamma$, to $\pm
1.5$\,K\,km$^{-1}$. If the accuracy of each temperature
measurement is better, say to $\pm 1$\,K, then the corresponding
accuracy on $\Gamma$ reduces to $\pm 0.8$\,K. Taking these constraints
on the lapse rate in the standard AOS atmosphere,
i.e.\ $\Gamma_\rmn{T} = -7.28\pm 1.5$\,K\,km$^{-1}$, we plot the
relative change in \param in \prettyref{fig:profilerconstraints}. The
plot looks very similar to \prettyref{fig:groundTconstraints}. The
error on \param in the submillimetre windows, where the dispersion is
significant, is between 1 and 2\,per cent using the information from
the temperature profiler. If the accuracy of the profiler exceeds
specifications to measure the temperature to $\pm 1$\,K then this
constraint reduces to around 0.5--1\,per cent.

\section{Summary}

Both of the primary phase-calibration techniques for ALMA rely on
accurate estimates of how the atmospheric path fluctuations vary with
frequency, because:
\begin{enumerate}
\item The phase solutions need to be scaled from the calibration to
  the science frequency for phase-transfer fast-switching.
\item The WVRs essentially measure variations in the quantity of water
  vapour, which must be converted into fluctuations in the phase delay
  of the incoming signal.  
\end{enumerate}

First, we revisited the magnitude of the wet dispersive path delay
in a median AOS atmosphere using the \atm software:
\begin{itemize}
\item In Bands 3 and 4 (84--163\,GHz), the wet dispersive path delay is a small
  fraction, 0.5--3\,per cent, of the wet non-dispersive path delay.
\item For Bands 5 and above (163--720\,GHz), the wet dispersive path
  delay becomes a significant fraction of the non-dispersive, $\gtrsim
  5$\,per cent and should be considered in any analysis, particularly
  at the highest frequencies. In the worst case, Band 8 (385--500\,GHz), the
  dispersive path delay is 20--55\,per cent of the non-dispersive. 
\item The variation in the dispersive path delay \emph{across} the 4-GHz
  instantaneous bandwidth of a single observation is typically 2--5\,per cent but
  does rise to higher fractions, $\sim$13\,per cent, in Band 9. Thus,
  the capability of the ALMA correlator and software to apply phase corrections
  channel-by-channel will prove useful.
\end{itemize}

Next, we investigated how the the dispersive path delay changes when
the model parameters were varied, over ranges that represented the
typical and extreme atmospheric conditions measured above the ALMA site:
\begin{itemize}
\item The amount of atmospheric water vapour or equivalently the airmass does not affect \param,
  i.e.\ the dispersive path contribution to the fluctuations. 
\item The typical changes in the water vapour
  scale height and ground pressure do produce small changes ($\lesssim 2$\,per cent)
  to \param.
\item The dispersive path delay depends more strongly on the temperature profile
  of the atmosphere, particularly the air temperature at the ground. The 10 to 90 percentiles of the Chajnantor
  ground-temperature distribution cause variations in the dispersive \param
  of 7--9\,per cent from the median. Additionally, typical diurnal variations in temperature ($\pm 5$\,K)
  would produce similar changes of $\pm 5$\,per cent. This will be
  significant at frequencies where the dispersive path delay provides a major contribution to the total
  path delay ($\gtrsim 345$\,GHz). \param also depends on \tlr
  with the typical variation being some 2--3\,per cent.
\end{itemize} 

These results indicate that obtaining a ground-level air temperature
estimate for each antenna to an accuracy of about
$\pm 2$\,K will reduce uncertainties in the models of the dispersive phase
to a satisfactory level. In combination with lapse rate estimates,
our calculations suggest that the ancillary calibration instruments on
site should be able to constrain the dispersive terms to 1--2\,per
cent, on target for the current calibration budget.

The relatively small expected variation of dispersive path scaling
with the natural range of atmospheric conditions at the AOS is also
encouraging from the perspective of \emph{empirical} models, which
use the observed correlations between phase and WVR
measurements. Because the variation is small, it means that the
frequency-dependence of such empirical models will \emph{not\/} need
to be re-calibrated very often.

\section*{Acknowledgments}

This work is supported by the European Commission's Sixth Framework
Programme as part of the wider `Enhancement of Early ALMA Science'
project, managed by the European Southern Observatory (ESO). The
workpackage at the University of Cambridge is led by John Richer and
previously by Richard Hills. Further information on the Cambridge
effort can be found on our webpages: \url{http://www.mrao.cam.ac.uk/projects/alma/fp6/}.

\renewcommand{\bibname}{References}

\appendix

\section{Distributions of site-characterization parameters} \label{apdx:sitedata}

\subsection{Water vapour column} \label{sec:watervapourcolumn}

The quantity of water vapour above Chajnantor is not a
directly-observable quantity. Instead, we calculate it from the
measured atmospheric opacity using an appropriate scaling for the
site. NRAO has operated an automated tipping radiometer at 225\,GHz on
the Chajnantor plateau since 1995 and the data up to August 2004 are
collected online \citep{sitecharacterization}. We summarize the
various cumulative distributions for $\tau_{225}$ used in the ALMA
memo series and online in \prettyref{tab:taudistribution}.

\begin{table}
\caption{Percentiles of the cumulative $\tau_{225}$ distribution above Chajnantor
  presented in previous work.}
\label{tab:taudistribution}
\begin{tabular}{l l l l l}
\toprule
Percentile & $\tau_{225}$~$^1$ & $\tau_{225}$~$^2$ & $\tau_{225}$~$^3$ & $\tau_{225}$~$^4$  \\
\midrule
25 & 0.036 & 0.037 & 0.037 & 0.037 \\
50 & 0.061 & 0.061 & 0.062 & 0.060 \\
75 & 0.115 & 0.122 & 0.125 & 0.118 \\
\bottomrule
\end{tabular} \\
$^{1}$ Memo 334 \citep{ALMAMemo334}, spanning 04/95 -- 07/00.\\
$^{2}$ Memo 471 \citep{ALMAMemo471}, spanning 04/95 -- 08/01.\\
$^{3}$ Memo 512 \citep{ALMAMemo512}, spanning 04/95 -- 08/03.\\
$^{4}$ The latest publicly-available data
\citep{sitecharacterization}, spanning 04/95 -- 12/04.\\
\end{table} 

Various memos have also computed the conversion relation between
\taucso and the PWV column (e.g.\ \citealt{ALMAMemo271};
\citealt{ALMAMemo333}). We use the relation presented in
\citet{giovanelli01}:
\begin{equation} \tau_{225} = 0.0435(\rmn{PWV}/\rmn{mm})+0.0068 \rmn{,}\label{eqn:tauconversion}\end{equation}
which is derived for Chajnantor from a comparison of \taucso to the PWV derived from
a 183-GHz WVR and agrees well with radiosonde data for
PWV<3\,mm. Using the standard atmospheric parameters we presented in
\prettyref{tab:atmparameters}, we also calculated the following conversion from \atm:
\begin{equation} \tau_{225} = 0.0416(\rmn{PWV}/\rmn{mm})+0.0120 \rmn{.}\label{eqn:tauconversionatm}\end{equation}
This is similar to \prettyref{eqn:tauconversion} but we prefer the
experimentally-derived scaling as it has been tested more
thoroughly. In \prettyref{tab:cdistribution} we list the
percentiles of the PWV distribution which we use in this memo, derived
from the most recent cumulative \taucso distribution that is publicly
available \citep{sitecharacterization} using
\prettyref{eqn:tauconversion}.

\begin{table}
\caption{Percentiles of the cumulative $\tau_{225}$ and $c$
  distributions above Chajnantor used in this memo. $c$ has been calculated from
  \taucso using \prettyref{eqn:tauconversion}.}
\label{tab:cdistribution}
\begin{tabular}{l l l}
\toprule
Percentile & $\tau_{225}$~$^1$ & $c$ (mm)  \\
\midrule
10 & 0.026 & 0.44 \\
25 & 0.037 & 0.69 \\
50 & 0.060 & 1.22 \\
75 & 0.118 & 2.56 \\
90 & 0.244 & 5.45 \\
\bottomrule
\end{tabular} \\
$^{1}$ Latest publicly-available site characterization data
\citep{sitecharacterization}, spanning 04/95 -- 12/04.\\
\end{table} 
 
\subsection{Parameters of the atmospheric vertical profile} \label{sec:verticalparams}

To derive distributions of atmospheric parameters which depend on
the variation of water vapour pressure and temperature with height
through the atmosphere, we make use of the library of radiosonde
launch data above the Chajnantor plateau. These radiosonde flights
were jointly operated by Cornell University, NRAO, ESO and the
Smithsonian Astrophysical Observatory between October 1998 and
December 2001. The data and an analysis of each dataset are
available on dedicated web
pages\footnote{\url{http://www.tuc.nrao.edu/alma/site/Chajnantor/instruments/radiosonde}}. A
preliminary analysis of these data was presented in
\citet{giovanelli01}. They show the vertical distribution of
water vapour density in a median atmosphere derived from 108
launches is well approximated by an exponential with scale height $h_0
= 1.135$\,km. We present the cumulative distribution of $h_0$ in
\prettyref{tab:h0distribution}. This is constructed using the analyses
of Bryan Butler, which fitted an exponential to the water vapour pressure between 0 and 10\,km above the
Chajnantor plateau for each radiosonde launch. We include 194 separate launches over varying months and times
of day.  

\begin{table}
\caption{Percentiles of the cumulative $h_0$ distribution used in this
  memo, constructed by Bryan Butler who fitted an exponential function
  to the data between 0 and 10\,km
  above Chajnantor (see \url{http://www.tuc.nrao.edu/alma/site/Chajnantor/instruments/radiosonde}).}
\label{tab:h0distribution}
\begin{tabular}{l l }
\toprule
Percentile & $h_0$ (km)  \\
\midrule
10 & 0.97 \\
25 & 1.06\\
50 & 1.16\\
75 & 1.29\\
90 & 1.54\\
\bottomrule
\end{tabular} \\
\end{table} 

Furthermore, to ascertain the distribution of the tropospheric lapse rate, \tlr
(\prettyref{tab:tlrdistribution}), we fit a straight line to the
radiosonde temperature data from 204 launches using a least-squares method. As most of the water
vapour is concentrated in the first layers of the atmosphere we fit
only to data in the first 1\,km above the plateau surface. We note
that in reality few temperature profiles from radiosonde data are a
perfect straight line and some show more complicated features such as
temperature inversions. However, for the purposes of this memo a simple
fit should be sufficient.   

\begin{table}
\caption{Percentiles of the cumulative \tlr distribution above
  Chajnantor used in this
  memo. It is constructed from fits to the temperature data $\leq
  1$\,km from the Earth's surface at Chajnantor (see \url{http://www.tuc.nrao.edu/alma/site/Chajnantor/instruments/radiosonde}).}
\label{tab:tlrdistribution}
\begin{tabular}{l l }
\toprule
Percentile & \tlr (K\,km$^{-1}$)  \\
\midrule
10 & $-9.71$\\
25 & $-8.83$ \\
50 & $-7.28$\\
75 & $-5.69$\\
90 & $-4.80$\\
\bottomrule
\end{tabular} \\
\end{table} 

\label{lastpage}

\end{document}